\documentstyle[twoside,fleqn,espcrc2]{article}

\def\pdot {\dot P}
\def\ltsima{$\; \buildrel < \over \sim \;$}
\def\lsim{\lower.5ex\hbox{\ltsima}}
\def\gtsima{$\; \buildrel > \over \sim \;$}
\def\gsim{\lower.5ex\hbox{\gtsima}}
\def\approxlt{\mathrel{\spose{\lower 3pt\hbox{$\sim$}}
        \raise 2.0pt\hbox{$<$}}}
\def\approxgt{\mathrel{\spose{\lower 3pt\hbox{$\sim$}}
        \raise 2.0pt\hbox{$>$}}}
\newcommand{\be}{\begin{equation}}
\newcommand{\en}{\end{equation}}

\newcommand{\AmS}{{\protect\the\textfont2
  A\kern-.1667em\lower.5ex\hbox{M}\kern-.125emS}}

\title{Recent Results on the Anomalous X--ray Pulsars}

\author{S. Mereghetti\address{Istituto di Fisica Cosmica del C.N.R., 
        v.Bassini 15, Milano, Italy}, %
        L. Stella$^{b}$ 
        and 
        G.L. Israel\address{Osservatorio Astronomico di Roma, 
        v. dell'Osservatorio 2, Monteporzio Catone, Italy}}
       
\input psfig.tex
\begin{document}

\begin{abstract}
The ''Anomalous X--ray Pulsars'' (AXP) are a small group of X--ray pulsars
characterized by periods in the $\sim$5-10 s range and by the absence
of massive companion stars.
There are now 7 possible members of this class of objects.
We review recent observational results on their X--ray spectra, 
spin period evolution, and searches for orbital motion  
and discuss the implications for possible models.

\end{abstract}

\maketitle

\section{THE ACCRETING X--RAY PULSARS}

There is compelling evidence that most of the
$\gsim$ 50 accreting pulsars
in X--ray binaries have massive companion stars.
They are therefore classified as High Mass X--ray Binaries (HMXRB). 
In most cases this is based on the observation of their optical
counterparts.
Ten X--ray pulsars  are optically identified with   OB supergiant stars,
while Be companions have been found for $\sim$20 pulsars.
Many of the optically unidentified X--ray pulsars 
are inferred to be HMXRB, because they share the properties of
the Be/neutron star systems (hard spectrum, transient behaviour, often 
with a recurrence equal to the orbital period).
 
The four   X--ray pulsars   optically identified with low mass
stars (Her X--1, GX1+4, GRO~J1744-28 and  4U~1626--67)
form a very inhomogenous group, each source
having its own peculiar properties \cite{Nag,Kov}.



In the last few years there has been growing evidence that a
group of   pulsars, clearly   not belonging to the HMXRB class,
possess the following prominent common properties \cite{MS}: 

- their optical counterparts are very faint, implying that 
they cannot have massive companions

- their spin periods are distributed in a narrow range ($\sim$5-10 s),
contrary to those of   HMXRB  that cover a much larger interval
 
- they have    very soft   X--ray spectra 
(with the exception of 4U~1626--67)

- their  X--ray luminosity is relatively low   
($10^{35}$-10$^{36}$ erg s$^{-1}$)
compared to that of HMXRB pulsars

- their X--ray flux shows little variability on timescales
from months to years (they are not transient systems)

- they have a relatively stable spin period evolution,
with long intervals of nearly constant spin-down

- a few of them are possibly associated to supernova remnants.

These objects have been called to in different ways by various
authors: "6 second pulsars" \cite{VTH}, 
braking pulsars \cite{GAW}, very low mass binary pulsars \cite{MS},
anomalous X--ray pulsars (AXP). 
In the following we shall 
adopt the latter name.

\begin{table*}[hbt]
\setlength{\tabcolsep}{1.5pc}
\newlength{\digitwidth} \settowidth{\digitwidth}{\rm 0}
\catcode`?=\active \def?{\kern\digitwidth}
\caption{AXP with two component spectra}
\label{tab:effluents}
\begin{tabular*}{\textwidth}{@{}l@{\extracolsep{\fill}}rcccc}
\hline
                 & \multicolumn{1}{r}{$kT$} 
                 & \multicolumn{1}{r}{$R_{bb}$} 
                 & \multicolumn{1}{r}{$L_{bb}$/$L_{tot}$} 
                 & \multicolumn{1}{r}{power law}         
                 & \multicolumn{1}{r}{ref.}         \\
                 & \multicolumn{1}{r}{(keV)} 
                 & \multicolumn{1}{r}{(km)} 
                 & \multicolumn{1}{r}{ } 
                 & \multicolumn{1}{r}{photon index }         
                 & \multicolumn{1}{r}{ }         \\
\hline
4U~0142+61            & $ 0.39$ & $2.4~d_{1kpc}$  & $  0.4  $ & $  3.7$  & \cite{W96} \\
1E~2259+586           & $ 0.44$ & $3.3~d_{4kpc}$  & $  0.4  $ & $  3.9$  & \cite{P98} \\
1E~1048.1-5937        & $ 0.64$ & $0.6~d_{3kpc}$  & $  0.5  $ & $  2.5$  & \cite{O98} \\
1RXS~J170849.0-400910 & $ 0.41$ & $4.5~d_{5kpc}$  & $  0.2  $ & $  2.9$  & \cite{S97} \\   
4U~1626--67         &$ 0.3-0.6$ & $3-5~d_{3kpc}$  & $ 0.1-0.2 $ & $  0.8$  & \cite{A95,Orl98} \\
 \hline
\end{tabular*}
\end{table*}

\section{AXP INVENTORY}

The 5 sources originally considered \cite{MS} in the AXP group are :
1E~2259+586,
1E~1048.1--5937,
4U~1626--67,
4U~0142+61, and
RXJ~1838.4--0301.

Thanks to recent ASCA discoveries,
two  new possible  members   have been   proposed.
An 11 s periodicity has been found \cite{S97} in the
source 1RXS~J170849.0-400910 first discovered during the ROSAT All Sky Survey.
Its very soft spectrum is similar to that of the AXP. 
Nothing is yet known on its   optical counterparts.
The second possible AXP is the unresolved X--ray source 1E~1841--045  
discovered with the Einstein satellite \cite{K85} near the geometrical center of
the young ($\sim2000$ yr) supernova remnant Kes 73 and recently found to be 
pulsed at 11.8 s \cite{VG}. 

Optical observations of the field of RXJ~1838.4--0301 revealed the presence
of a  main sequence K5  star with V$\sim14.5$ \cite{MBN}.
This star could be responsible for the observed X--ray flux, 
since the implied X--ray to optical flux ratio
is compatible with the level of coronal emission expected  
in late type stars. In the lack of an independent confirmation
of the 5.45 s pulsation \cite{S}  the inclusion of  RXJ~1838.4--0301  in the AXP
group should be considered tentative.

Another questionable AXP is 4U~1626--67. Though originally included 
in the AXP group \cite{MS},     various authors pointed out its 
different nature \cite{VTH,GAW}, on the basis of its hard spectrum and
of the fact that it is the only source of this group with an
identified optical counterpart and a  
well established binary nature.

\section{X--RAY SPECTRA}

Recent ASCA \cite{W96} and BeppoSAX \cite{P98,O98}
observations have shown that the
spectra of 1E~2259+586, 1E~1048.1-5937
and  4U~0142+61 cannot be described by single power laws.
Their spectra are well fitted by the combination of  a 
blackbody-like component with kT $\sim0.5$ keV and a steep 
power law with photon index $\alpha\sim$3-4 (see Table 1).
The same two-component spectrum is compatible with the ASCA data
for 1RXS~J170849.0-400910,
although a simple power law   gives  a formally acceptable fit \cite{S97}.

The spectrum of 1E~1841--045 \cite{VG} can be fit with a highly 
absorbed, soft power law ($\alpha\sim3$). 
A blackbody component would be difficult to detect
due to the high column density  (N$_{H}\sim2~10^{22}$
cm$^{-2}$).
RXJ~1838.4--0301 has been observed so far only in the ROSAT band (0.1-2.4 keV),
where it has a very soft ($\alpha\sim$3) though poorly constrained spectrum \cite{S}.

The spectrum of  4U~1626--67, 
a flat power law   followed by an
exponential cut-off above $\sim20$ keV \cite{Pr79},
is more typical of accreting pulsars in HMXRB. 
However, the presence of a soft blackbody component
has been reported by several authors \cite{A95,Orl98}. 
A complex spectral feature around 1 keV,
interpreted as emission from hydrogen-like neon \cite{A95},
might indicate that the companion star is an He burning star.

\section{PERIOD EVOLUTION}

Spin-down on a timescale of $\sim$10$^{4}$--10$^{5}$ years is one of the
distinctive properties of AXP (see Table 2).
Variations in the spin-down rate have been observed  in the sources
for which many period measurements are available \cite{Me96,P98,O98}.
In the case of 1E~2259+586, 
these fluctuations are consistent with the torque noise measured
in accreting systems \cite{BS} (that is several orders of magnitude
greater than that of radio pulsars).

4U~1626--67 again stands out for its different properties. 
In fact this source has been nearly steadily spinning-up 
for more than a decade after its discovery.
In 1990 it underwent a rapid torque reversal and,
since then, it has been spinning down at a rate nearly equal
to the previous spin-up rate \cite{Ch97}.

Nothing is known yet on the period evolution of 1RXS~J170849.0-400910 and RXJ~1838.4-0301.

\section{SEARCHES FOR ORBITAL MOTION}

No periodic X--ray flux modulations such as dips
or eclipses have been detected in AXP.
In the lack of optical identifications, 
the search for Doppler delays in the pulse arrival times
is the only way to assess their possible binary nature.
This has been done with different satellites 
for the sources listed in Table 2,
where the upper limits on   $a_x\sin i$  are summarized.

The most constraining limits    are those
recently derived  with the RossiXTE satellite for 1E~1048.1-5937 and 1E~2259+586 \cite{MIS}, as well as the value of 8 light-ms for  4U 1626--67 \cite{Shi}.

In this source, in addition to
the X--ray  periodicity at 7.7~s, a pulsation at a
slightly lower  frequency is present in the optical band \cite{Mid}. This is  probably due to reprocessing of the X--ray pulses occurring
near the  companion star, and the difference of the two periodicities can be 
explained with an orbital period of 41.4~min. 
Since this orbital period has not been confirmed  
with independent methods, we give in Table 2 also the limits
on   $a_x\sin i$ for other possible values of $P_{orb}$.

\begin{table*}[hbt]
\setlength{\tabcolsep}{1.pc}
\caption{AXP Timing properties}
\label{tab:effluents}
\begin{tabular*}{\textwidth}{@{}l@{\extracolsep{\fill}}rrlcc}
\hline
                 & \multicolumn{1}{r}{$P$} 
                 & \multicolumn{1}{r}{$P$/$\pdot$} 
                 & \multicolumn{1}{r}{$a_x\sin i$} 
                 & \multicolumn{1}{r}{Range of $P_{orb}$} 
                 & \multicolumn{1}{r}{ref.}   \\
                 & \multicolumn{1}{r}{(sec)} 
                 & \multicolumn{1}{r}{(years)} 
                 & \multicolumn{1}{r}{(light-sec)} 
                 & \multicolumn{1}{r}{  } 
                 & \multicolumn{1}{r}{  }   \\
\hline
4U~0142+61            &$  8.69 $&$ (4-15)~10^{4} $&$ 0.37~(99\%)    $& 430 s - 12 hr      & \cite{IMS}   \\
1E~2259+586           &$  6.98 $&$ 4~10^{5}      $&$ 0.03~(99\%)    $& 194 s - 1 day      & \cite{MIS}   \\
1E~1048.1-5937        &$  6.45 $&$ (5-14)~10^{3} $&$ 0.06~(99\%)    $& 200 s - 1 day      & \cite{MIS}   \\
4U~1626--67           &$  7.66 $&$ 5~10^{3}      $&$ 0.008~(3\sigma)$& $P_{orb}$ = 42 min & \cite{Shi}   \\
                      &$       $&$               $&$ 0.013~(3\sigma)$& 600 s - 10 hr      & \cite{Lev88}   \\
                      &$       $&$               $&$ 0.1~(2\sigma)  $&  1 d - 2 d         & \cite{J78}   \\
                      &$       $&$               $&$ 0.06~(2\sigma) $&  2 d - 60 d        & \cite{Ch97}   \\
                      &$       $&$               $&$ 0.15~(2\sigma) $&  60 d - 900 d      & \cite{Ch97}   \\
1RXSJ170849-400910    &$ 11.00 $&$     -         $&$   -         $&$                  $& \cite{S97}   \\     
1E~1841--045          &$ 11.77 $&$ 8~10^{3}      $&$   -         $&$                  $& \cite{VG}   \\
RXJ~1838.4--0301      &$  5.45 $&$     -         $&$   -         $&$                  $& \cite{S}   \\
\hline
\end{tabular*}
\end{table*}

%
\begin{figure}
\centerline{\psfig{figure=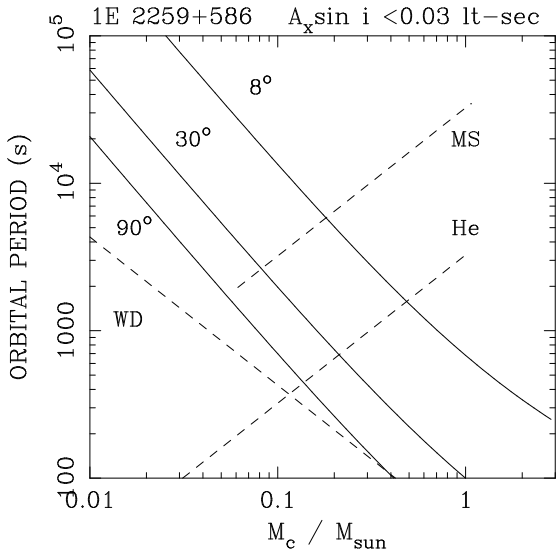,width=6cm,height=7cm} }
\caption{
Orbital onstraints from the  $a_x\sin i$ limit   for    1E~2259+586.}
\label{fig:toosmall}
\end{figure}

\section{DISCUSSION}

Though the precise nature of the low mass companion of 4U~1626--67
is still unclear, it is well established that 
this source consists of a neutron star accreting in  a binary system.
Considering the differences discussed above between 4U~1626--67 and the 
other sources, it is not straightforward to assume a similar model
for all the AXP.
Though an optical counterpart as faint as that of 4U~1626--67 
(Mv $\sim5$) cannot be ruled out  at the highly 
absorbed positions of some of  the other AXP,
a more serious inconsistency is the difference in spectral shape.
It is unlikely that this results from 
a different inclination  
of the line of sight, considering the similarities among the spectra of HMXRB pulsars
that likely encompass systems at different orbital inclinations.

The   $P$ and $\pdot$ values measured  in AXP imply a rotational energy loss
of the order of $10^{32}-10^{33}$ erg s$^{-1}$ (in the case of neutron stars with $I$ = 10$^{45}$ g cm$^{2}$),
too small to power the observed luminosities.
A spinning down, isolated white dwarf would provide a larger
rotational energy loss thanks to its higher momentum of inertia.
Such a model was originally proposed \cite{M88,Pa90} for
1E~2259+586.
However this possibility has been ruled out by the $\pdot$ variations
observed in the last few years (see section 3).
Indeed the observed variations in  $\pdot$, correlated with luminosity 
changes and occasionally implying short spin-up intervals,
are a strong indication that accretion is occurring in these systems
\cite{Me96,BS}.

%

It has been proposed \cite{VTH} that the AXP (with the exception of 4U~1626--67)
are the descendant of Thorne-Zytkow objects and consist of isolated
neutron stars accreting from residual disks.
The blackbody component observed in some AXP has been interpreted as 
evidence for quasi-spherical accretion onto an isolated neutron 
star formed after common envelope evolution and
spiral-in of a massive X-ray binary \cite{W96,GAW}.  
In this case, the accretion flow results from the remaining part of
the  massive star envelope and  is thought to consist of two components:
a low-angular momentum component, giving rise to the
blackbody emission from a considerable fraction of the neutron star surface,
and a high-angular momentum one, forming an accretion disk responsible
for the power-law emission and for the long term
spin-down evolution. 

Though this is certainly a very interesting interpretation to be
investigated with future observations, a more standard binary scenario
is also possible. 
In fact,  the presence of very low mass white dwarfs or He-burning companions
cannot be ruled out, despite the   limits on  $a_x\sin i$ .

In   the case of 1E~2259+586
this is illustrated in figure 1, where the limits on orbital period, $P_{orb}$,   
versus mass of the companion, $M_c$,  are plotted 
assuming three different values for the unknown inclination angle.
The dashed lines indicate the positions of Roche-lobe filling companions
under the assumption of conservative mass transfer driven by angular 
momentum losses due only to gravitational radiation \cite{VH}.
They refer  to the cases of a  main sequence,  a 
He burning star and a fully degenerate hydrogen white dwarf.
Values of $P_{orb}$ and $M_c$ below the corresponding dashed line are excluded,
while those above the lines require accretion through stellar wind.

It is clear that, unless all these systems are nearly face-on,
the most likely mass donors are   white dwarfs and helium stars
with masses below a few tenths of solar masses.
In the latter case, 
the mass transfer is most likely through a stellar wind,
since the accretion expected from a Roche-lobe
filling He-star would produce a far greater luminosity than the 
observed one.

\end{document}